\documentclass[twocolumn,twocolappendix,tighten]{aastex63}
\usepackage{amssymb, amsfonts, amsmath,framed}
\usepackage{lineno}
\nolinenumbers

\usepackage{gensymb}
\usepackage{bm}
\usepackage{multirow}
\usepackage{mathrsfs}

\usepackage{graphicx}
\usepackage{stackengine}
\usepackage[caption=false]{subfig}
\usepackage{threeparttable}
\usepackage{tabularx}

\usepackage {verbatim}

\usepackage{float}

\expandafter\ifx\csname package@font\endcsname\relax\else
 \expandafter\expandafter
 \expandafter\usepackage
 \expandafter\expandafter
 \expandafter{\csname package@font\endcsname}
\fi
\expandafter\ifx\csname package@font\endcsname\relax\else
 \expandafter\expandafter
 \expandafter\usepackage
 \expandafter\expandafter
 \expandafter{\csname package@font\endcsname}
\fi
\newcommand{\be}{\begin{equation}}
\newcommand{\ee}{\end{equation}}
\newcommand{\bi}{\begin{itemize}}
\newcommand{\ei}{\end{itemize}}
\newcommand{\figg}[1]{Fig.~\ref{fig:#1}}

\newcommand{\eq}[1]{Eq.~\ref{eq:#1}}

\newcommand{\ls}{\textcolor{red}}
\newcommand{\fix}[1]{\textcolor{red}{[fix] }}

\newcommand{\gammarad}{\gamma_{\rm rad}}
\newcommand{\gammacool}{\gamma_{\rm cool}}

\newcommand{\tacc}{t_{\rm  acc}}
\newcommand{\tesc}{t_{\rm  esc}}
\newcommand{\omcm}{\omega_{\rm  c}^{-1}}

\begin{document}
\title{The effect of inverse Compton losses on particle acceleration \\ in three-dimensional relativistic reconnection}

\author[0009-0000-1347-2419]{Ian Bowyer}
\email{ibowyer@purdue.edu}
\affiliation{Department of Physics and Astronomy, Purdue University, West Lafayette, IN, 47907, USA}
\author[0000-0003-1503-2446]{Dimitrios Giannios}
\affiliation{Department of Physics and Astronomy, Purdue University, West Lafayette, IN, 47907, USA}
\author[0000-0002-1227-2754]{Lorenzo Sironi}
\affiliation{Department of Astronomy and Columbia Astrophysics Laboratory, Columbia University, New York, NY, 10027, USA}
\affiliation{Center for Computational Astrophysics, Flatiron Institute, 162 5th Avenue, New York, NY, 10010, USA}

\date{\today}

\begin{abstract}
Relativistic magnetic reconnection is a key mechanism for dissipating magnetic energy and accelerating particles in astrophysics. In the absence of radiative cooling, recent particle-in-cell (PIC) simulations have shown that high-energy particles gain most of their energy in the upstream region, during a short-lived ``free phase'' where they meander between the two sides of the layer; when they get captured/trapped by the downstream flux ropes, they undergo a ``trapped phase,'' where no significant energization occurs.
Here, we perform a suite of 3D PIC simulations of relativistic reconnection including inverse Compton (IC) losses in the weakly cooled regime in which the radiation-reaction-limited Lorentz factor $\gamma_{\rm rad}$ exceeds the magnetization $\sigma$. We show that electron cooling losses do not appreciably alter the reconnection rate, the structure of the layer, and the physics of particle acceleration in the free phase, so the spectrum of free electrons is $dN_{\rm free}/d\gamma\propto \gamma^{-1}$, as in the uncooled case. The spectrum of trapped electrons above the cooling break $\gamma_{\rm cool}$ (in the range $\gamma_{\rm cool}<\gamma<\gamma_{\rm rad}$) is $dN/d\gamma\propto \gamma^{-3}$, steeper than the scaling  $dN/d\gamma\propto \gamma^{-2}$ of uncooled simulations. This confirms that no significant particle energization occurs during the trapped phase. Our results validate the model by \citet{Zhang_23ads} for particle acceleration in 3D relativistic reconnection, and imply that radiative emission models of reconnection-powered astrophysical sources should employ a two-zone structure, that differentiates between free, rapidly accelerating particles and trapped, passively cooling particles.  
\end{abstract}


%
%

\section{Introduction}
\label{sec:introduction}

Relativistic magnetic reconnection---where the magnetic enthalpy density is greater than the plasma relativistic enthalpy density \citep[for a review, see][]{annurev_DLU}---is an important mechanism for dissipating magnetic energy and accelerating particles to relativistic energies in astrophysical plasmas \citep{lyutikov_uzdensky_03,lyubarsky_05,giannios_09,giannios_13,comisso_14}, especially in the magnetospheres of compact objects and in relativistic jets. It is invoked as a source of fast, bright, nonthermal flares, and possibly as the engine producing Ultra High Energy Cosmic Rays \citep[UHECRs; e.g.][]{zhang_sironi_21,statho_24} and  high-energy neutrinos \citep{fiorillo_24,karavola_25,karavola_25b}.

In recent years, our understanding of particle acceleration in relativistic reconnection has significantly advanced, thanks to large-scale, first-principles particle-in-cell (PIC) simulations \citep{annurev_DLU}. Most advances have come from electron-positron runs, which are computationally cheaper than electron-ion runs. In pair plasmas, 
the particle energy spectrum of the reconnected plasma is modeled as a broken power law, with break Lorentz factor $\gamma_{\rm br}\sim\sigma$ \citep{werner_16,comisso_23}. Here, $\sigma$ is the magnetization, i.e., the ratio of magnetic enthalpy density and rest mass energy density for the cold upstream plasma. 

For particles in the low-energy range ($\gamma\lesssim \gamma_{\rm br}$), acceleration is  a rapid, one-shot, ``injection'' process, which boosts them from the low, non-relativistic upstream energies up to ultra-relativistic energies. While the injection physics (i.e., acceleration up to $\sim \gamma_{\rm br}$) is similar in 2D and 3D \citep[e.g.,][]{sironi_22,french_23,totorica_23,french_25,gupta_25}, the process of particle acceleration to energies beyond the spectral break differs dramatically between 2D and~3D. In 2D, the highest-energy particles are trapped in plasmoids \citep{sironi_spitkovsky_14,guo_14,werner_16,uzdensky_22,petropoulou_18,hakobyan_21}, i.e., in the magnetic islands / flux ropes generated by reconnection. 

In 3D, the flux-rope kink instability breaks the invariance of magnetic flux ropes along the $z$-direction of the electric current, allowing some particles to escape back upstream by moving along~$z$. 
Escaping particles with Lorentz factors $\gamma\gtrsim \,\sigma$ have Larmor radii large enough that they can sample both upstream sides of the reconnection layer and get efficiently accelerated \citep{zhang_sironi_21}. The acceleration process can be described in two equivalent ways: an energetic particle crosses the current sheet from one upstream region to the other, completing an arch-shaped segment of its cyclotron orbit in each region. Since the field reverses, these arches always return the particle back to the layer, and its $z$-displacement is always in the same direction, which allows the particle to move nearly along the upstream motional electric field $E_{\rm rec}\sim\eta_{\rm rec} B_0$, where $B_0$ is the upstream field strength and $\eta_{\rm rec}\sim 0.1$ is the reconnection rate for a collisionless relativistic plasma \citep[e.g.,][]{selvi_23,moran_25,ripperda_26}. Equivalently, the particle is confined in between the two converging upstream flows, and gets energized via a Fermi-like process \citep{giannios_10}. In doing so, the particle accelerates continuously at nearly the maximum rate, with $\dot{\gamma}_{\rm acc}\sim \eta_{\rm rec}\omega_{\rm c}$, where $\omega_{\rm c}=eB_0/mc$ is the cyclotron frequency \citep[][]{speiser_65,uzdensky_11,cerutti_12a,zhang_sironi_21}.

Motivated by the investigation of particle  orbits in \citet{zhang_sironi_21}, we proposed a model for the formation of the power-law distribution at $\gamma\gtrsim \gamma_{\rm br}$ in 3D \citep{Zhang_23ads}, which is valid for weak guide fields (the guide field is the non-reversing field component along $z$).
It assumes that high-energy particles gain most of their energy in the upstream region, during a short-lived ``free phase'' where they meander between the two sides of the layer, as described above. The acceleration time is $t_{\rm acc}\sim \gamma/ \eta_{\rm rec}\omega_{\rm c}$.
They leave the region of active acceleration after a time $t_{\rm esc}$, when they get captured/trapped by the downstream flux ropes; during the subsequent ``trapped phase,'' no significant energization occurs. 3D PIC simulations \citep{Zhang_23ads} show that $t_{\rm esc}\simeq t_{\rm acc}$ for magnetizations $\sigma\gtrsim {\rm few}$, which leads to a universal (i.e., nearly $\sigma$-independent) power-law spectrum $dN_{\rm free}/d\gamma\propto \gamma^{-1}$ for the free particles undergoing active acceleration. The spectrum of trapped particles---which dominate the overall particle count, since the free phase is extremely short-lived---can be obtained by assuming a quasi-steady state: at each energy, the rate of free particles getting trapped (after $t_{\rm esc}\propto \gamma$) equals the rate of trapped particles leaving the system (on $t_{\rm adv}\sim L/c$, where $L$ is the half-length in the outflow direction), yielding $dN/d\gamma\propto (t_{\rm adv}/t_{\rm esc})\; dN_{\rm free}/d\gamma \propto \gamma^{-2}$, for all $\sigma\gtrsim{\rm few}$. 
 
In this paper, we introduce inverse Compton (IC) cooling losses in 3D simulations of relativistic reconnection, with the goal of validating the power-law formation model proposed by \citet{Zhang_23ads}. The physics of 3D relativistic reconnection in the presence of cooling losses is still relatively under-explored, for both synchrotron \citep{Cerutti_2014,Chernoglazov_23ads,schoeffler_23} and IC cooling \citep{sironi_beloborodov_20}. In contrast to the case of synchrotron cooling, where losses depend both on particle energy and pitch angle, the case of IC cooling, where losses are only a function of particle energy, provides a simpler, more easily interpretable test of the acceleration physics discussed in \citet{zhang_sironi_21} and \citet{Zhang_23ads}. In this paper, we perform a suite of 3D PIC simulations of relativistic reconnection including IC losses in the weakly cooled regime in which the radiation-reaction-limited Lorentz factor $\gamma_{\rm rad}$ exceeds the magnetization \citep{annurev_DLU}. We show that electron cooling losses do not appreciably alter the reconnection rate, the structure of the layer, and the physics of particle acceleration in the free phase, so the spectrum of free electrons is $dN_{\rm free}/d\gamma\propto \gamma^{-1}$, as in the uncooled case. The spectrum of trapped electrons above the cooling break $\gamma_{\rm cool}$ (in the range $\gamma_{\rm cool}<\gamma<\gamma_{\rm rad}$) is $dN/d\gamma\propto \gamma^{-3}$, i.e., steeper than in the uncooled case. This lends support to one of the fundamental assumptions of the model by  \citet{Zhang_23ads}, i.e., that no significant energization occurs in the trapped phase.

 This paper is organized as follows. Section \ref{sec:2modelstuff} expands the analytical model by \citet{Zhang_23ads} to include the effect of cooling losses. 
In Section \ref{sec:3simulationsetup} we describe the PIC simulation setup and in Section \ref{sec:4results} we present the results of our simulations, and compare them with the analytical expectations presented in Section \ref{sec:2modelstuff}. In Section \ref{sec:5discussion} we discuss the astrophysical implications of our results, before concluding in Section  \ref{sec:6conclusion}.

\section{Particle Acceleration in Radiative Reconnection}
\label{sec:2modelstuff}
In this section, we start by defining the relevant energy scales (or equivalently, particle Lorentz factors) in radiative reconnection, and then we derive the expected form of the free and trapped particle spectra, for both uncooled and cooled cases.

\subsection{Energy Scales}
\label{sec:22energyscales}
We first summarize the main energy scales of radiative reconnection, assuming that radiation effects can be treated in the classical (i.e., non quantum) limit. The magnetization quantifies the mean energy per particle,
    \begin{equation}\label{eq:1mag}
    \sigma=\frac{B_0^2/4 \pi}{n_0 mc^2}~,
    \end{equation}
where $B_0$ and $n_0$ are the magnetic field and number density in the upstream, and we have adopted a cold pair-plasma composition with particle mass $m$.

In the absence of cooling losses, the Lorentz factor of the highest energy particles, $\gamma_{\rm  max}$, is obtained by balancing the acceleration time $t_{\rm acc}$ with the advection time out of the layer $t_{\rm adv}$, yielding  
    \begin{equation}\label{eq:2gammax}
    \gamma_{\rm  max}=\frac{eE_{\rm rec} L}{m c^2}~,
    \end{equation}
where $E_{\rm rec}\sim \eta_{\rm rec} B_0$ is the electric field and $L$ the half-length of the system along the outflow direction. 

In the presence of cooling losses---here, IC losses in the Thomson regime---we can define the radiation-reaction-limited Lorentz factor $\gamma_{\rm rad}$ at which the radiation-reaction drag force balances the accelerating force from the reconnection electric field, yielding
    \begin{equation}\label{eq:4gamrad}
    \gamma_{\rm  rad} = \sqrt{\frac{3 e E_{\rm rec}}{4 \sigma_{\rm T} U_{\rm ph}}} ~,
    \end{equation}
where $U_{\rm ph}$ is the photon energy density and $\sigma_{\rm T}$ the Thomson scattering cross section. We define cooling as strong when $\gamma_{\rm rad} < \sigma$, and weak in the opposite case.
A break  in the particle spectrum is expected at the cooling Lorentz factor $\gammacool$ at which the cooling time is comparable to the advection time, which yields 
    \begin{equation}\label{eq:5gamcool}
    \gamma_{\rm  cool} = \frac{3mc^2}{4 \sigma_{\rm T}  U_{\rm ph} L}=\frac{\gammarad^2}{\gamma_{\rm max}}~.
    \end{equation}
We define the fast-cooling regime as $\gamma_{\rm cool}<\sigma$ and the slow-cooling regime in the opposite case.

\subsection{Free Particles}
\label{sec:24free}
As discussed above, in 3D relativistic reconnection the presence of free particles (i.e., particles in the free phase of active acceleration) is most evident for Lorentz factors $\gamma>\sigma$, so we focus on this energy range and assume that cooling is weak, $\gammarad\gg \sigma$. The distribution of free particles in steady state can be described with a Fokker-Planck approach, where the equation for $f_{\rm  free}=d N_{\rm  free}/d\gamma$ reads
\begin{equation}\label{eq:6fp}
    \frac{\partial}{\partial \gamma} \left[(\dot{\gamma}_{\rm  acc}+\dot{\gamma}_{\rm  cool}) f_{\rm  free}\right] + \frac{f_{\rm  free}}{t_{\rm  esc}} = Q_{\rm  inj} \delta(\gamma-\gamma_{\rm  inj})~.
\end{equation}
Here, we have assumed that particles are injected into the free phase at $\gamma_{\rm inj}\sim \sigma$ with rate $Q_{\rm inj}$. As discussed by \citet{Zhang_23ads}, the results are the same if injection is not monoenergetic, as long as it drops sufficiently fast with $\gamma$. We define the acceleration time as $t_{\rm acc} =t_{\rm acc, 0}\gamma$, where $1/t_{\rm acc, 0}=\dot{\gamma}_{\rm acc}\sim \eta_{\rm rec}\omega_{\rm c}$. The escape time $t_{\rm esc}$ for free particles is the time after which they get trapped by the downstream flux ropes, thus terminating the free phase of active acceleration. 3D PIC simulations by \citet{Zhang_23ads} found that for $\sigma\gtrsim 3$, the escape time $t_{\rm esc}$ scales linearly with $\gamma$ and that $t_{\rm esc, 0}=t_{\rm esc}/\gamma\simeq t_{\rm acc, 0}$, i.e., free particles spend most of their life actively accelerating.

The cooling rate can be written as $\dot{\gamma}_{\rm cool}=-k \gamma^2$, where $k=\dot{\gamma}_{\rm acc}/\gammarad^2=\gammacool/t_{\rm adv}$. While we focus on IC cooling, the results of this section can also be applied to synchrotron cooling, as long as the free particles with $\gamma> \sigma$ have large pitch angles and primarily move in the (nearly uniform) upstream magnetic field. As shown by \citet{Zhang_23ads}, both assumptions are well satisfied.

The solution of \eq{6fp} for $\gamma_{\rm inj}<\gamma\ll \gammarad$ is 
\begin{equation}\label{eq:8ffree}
    f_{\rm  free}=Q_{\rm inj} \left(\frac{\gamma}{\gamma_{\rm inj}}\right)^{-s_{\rm free}}~,
\end{equation}
which is the same as in the uncooled case. Here, $s_{\rm free}={t_{\rm acc, 0}}/{t_{\rm esc, 0}}$; the numerical results by \citet{Zhang_23ads} suggest $s_{\rm free}\simeq 1$. In short, the distribution of free particles is expected to be the same as in the uncooled case, with the only difference that its cutoff is now determined by $\gammarad$ rather than $\gamma_{\rm max}$.

\subsection{Trapped Particles}
\label{sec:25trap}
A fraction of the particles flowing into the reconnection layer will never experience a free phase. They are mostly confined to low energies; in fact, \citet{Zhang_23ads} found that in uncooled simulations, most of the particles reaching $\gamma\gg \sigma$ have experienced at least one free phase. In this subsection, we focus on the trapped particles that have been free during part of their prior life. 
Even though the spectrum of free particles is confined to $\gamma>\gamma_{\rm inj}\sim \sigma$, as a result of cooling the spectrum of trapped particles can extend below $\sigma$. However, at such low energies the number of trapped particles (with a prior free phase) is much smaller than the number of particles that never experienced a free phase, which then control the spectral shape at $\gamma\lesssim \sigma$. Therefore, this section focuses on Lorentz factors $\gamma>\gamma_{\rm inj}\sim \sigma$.

One of the fundamental assumptions of the model by \citet{Zhang_23ads} is that trapped particles do not significantly energize. It follows that their distribution $f_{\rm  trap}=d N/d\gamma$ must obey
\begin{equation} \label{eq:11trapdiff}
    \frac{\partial}{\partial\gamma}\left(\dot{\gamma}_{\rm  cool}f_{\rm  trap} \right) + \frac{f_{\rm  trap}}{t_{\rm  adv}} = \frac{f_{\rm  free}}{t_{\rm  esc}}~,
\end{equation}
where we have assumed that the injection of trapped particles is dominated by free particles escaping (i.e., terminating) their free phase when captured by flux ropes. Trapped particles leave the system on the energy-independent timescale $t_{\rm adv}\sim L/c$. 

In the slow-cooling regime $\sigma\ll \gammacool$, the solution of \eq{11trapdiff} for $\sigma\ll \gamma\ll\gammacool$ is the same as in the uncooled case discussed by \citet{Zhang_23ads}, 
\begin{equation} \label{eq:13nocooldiff}
      {f_{\rm trap}} =\frac{{t_{\rm  adv}}}{t_{\rm esc, 0} \gamma}{f_{\rm  free}}\propto \gamma^{-s_{\rm free}-1}~.
\end{equation}
In contrast, beyond the cooling break, for  $\max[\gammacool,\sigma]\ll \gamma\ll\gammarad$, 
the solution is 
\begin{equation} \label{eq:14cooldiff}
      {f_{\rm trap}}  =\frac{\gammarad^2}{\gamma^2}{f_{\rm  free}}\propto \gamma^{-s_{\rm free}-2}~,
\end{equation}
which implies that $f_{\rm trap}\sim f_{\rm free}$ at $\gamma\sim \gammarad$.

\section{Simulation Setup}
\label{sec:3simulationsetup}
We perform 3D PIC simulations with TRISTAN-MP \citep{buneman_93, spitkovsky_05}. We use the Harris configuration to initialize the magnetic field and the current sheet, where the magnetic field of strength $B_0$  reverses from $+\hat{x}$ to $-\hat{x}$ across a current sheet at $y=0$.  
The field strength $B_0$ is parameterized by the magnetization $\sigma = B_0^2 / 4\pi  n_0 m c^2 = \left(\omega_{\rm  c} / \omega_{\rm  p}\right)^2$, where $\omega_{\rm  c} = e B_0 / m c$ is the gyrofrequency and $\omega_{\rm  p} = \sqrt{4\pi n_0 e^2 / m}$ is the plasma frequency. We choose $\sigma=10$ as a representative case of relativistic reconnection.
We consider an electron-positron composition, and the upstream plasma number density (including both species) is denoted by $n_0$ and modeled with two computational particles per cell. 
We also initialize a uniform ``guide'' field of strength $B_{\rm g}$ along $z$. Most of our runs employ $B_{\rm g}=0.1\, B_0$, but our results are nearly the same in the case of zero guide field, as we show in Appendix \ref{appendix:b}.    
We use periodic boundary conditions along the $z$ direction of the electric current (and of the guide field) and outflow boundary conditions in the $x$ direction of the reconnection exhausts. Along the inflow $y$ direction, two injectors continuously introduce fresh plasma and magnetic flux into the domain, while receding away from the midplane. This allows to reliably assess the properties of reconnection once the layer has achieved a statistical steady state.

We resolve the plasma skin depth $c/\omega_{\rm p}$ with 2.5 cells. All of the simulations have the same box size, with $L_z=L_x/2=L=800\,c/\omega_{\rm p}$. We vary the degree of IC cooling by changing $\gammarad$, and we explore  $\gammarad=30$, 43 and 60 (corresponding respectively to $\gammacool\simeq 3.5$, 7.2 and 14.2. We also compare our results with the uncooled case $\gammarad\rightarrow \infty$ discussed by \citet{Zhang_23ads}. 

As mentioned above, we adopt an electron-positron composition. We apply radiative cooling only to electrons, while positrons are uncooled. For an uncooled electron-ion plasma with a large ion magnetization, the post-reconnection scale separation between species reduces due
to efficient electron heating to ultra-relativistic temperatures (which then lowers the ratio of effective masses), so that relativistic electron-ion reconnection behaves
similarly as in pair plasma \citep{annurev_DLU}. This justifies our choice of neglecting the cooling losses of the positively charged particles, if our positrons are meant to represent heavier ions of the same (ultra-relativistic) energy. 
This choice has two other desirable consequences, given that our main goal is to test the power-law formation model by \citet{Zhang_23ads}, which was based on uncooled simulations. First, for every simulation, the positron properties in our cooled runs should match those of corresponding uncooled cases. Second, since the model by \citet{Zhang_23ads} was based on uncooled simulations, the most appropriate way to compare our results to their model would require that changes in the layer be kept  minimal. This can be achieved by ensuring that plasma pressure still appreciably contributes to force balance in the layer. If both species were to be cooled, this would only occur in the slow-cooling regime $\gammacool\gg \sigma$. In the case of uncooled positrons, their pressure still contributes, regardless of whether electrons are fast- or slow-cooling. 
In Appendix \ref{appendix:b}, we show that for $\gammarad=60$ (corresponding to $\gammacool\simeq 14.2\gtrsim \sigma$, so marginally slow cooling) our results are the same as in the case when both species are subject to radiative losses.

\begin{figure}[hbt!]
\centering
    \includegraphics[width=\columnwidth ] 
    {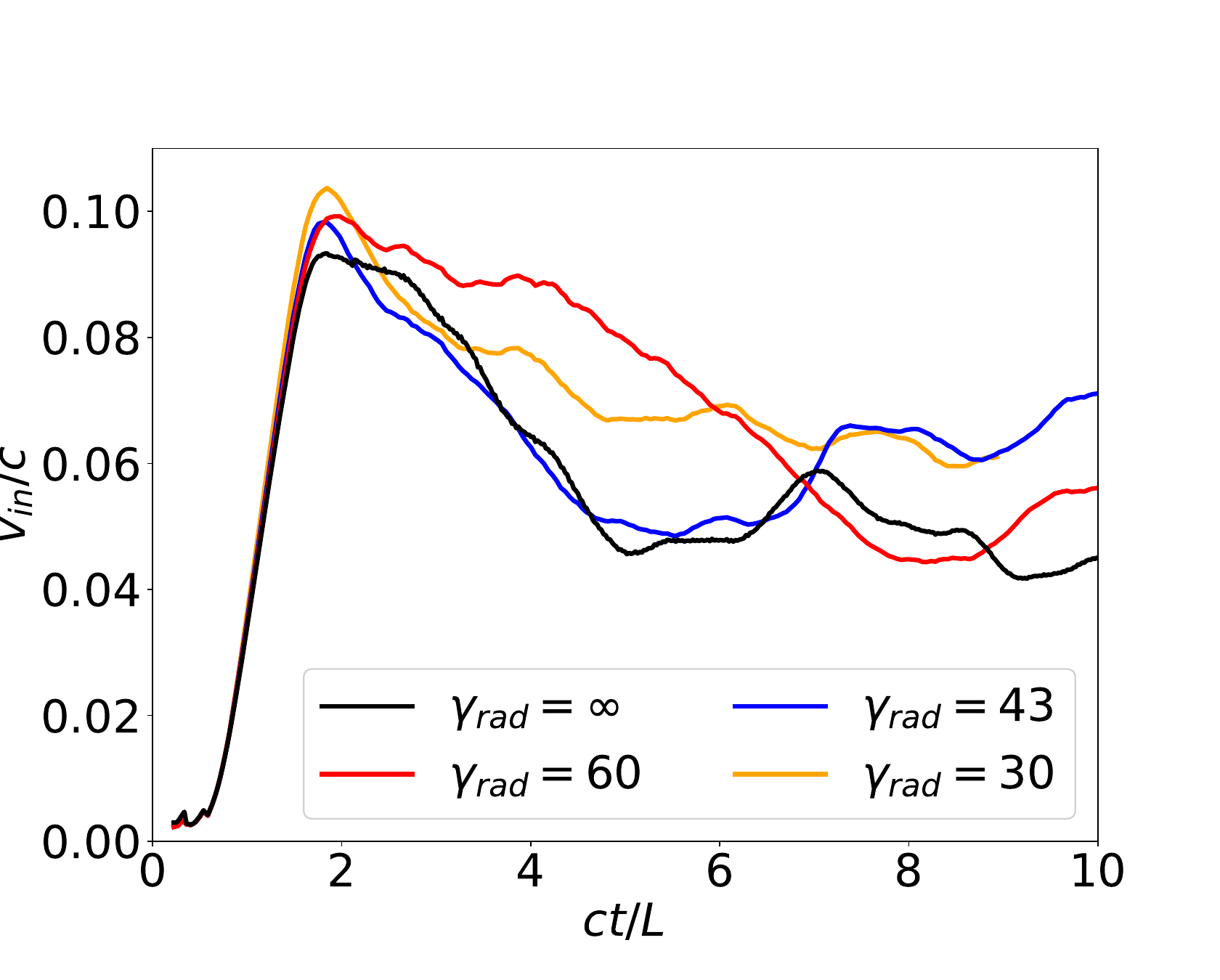}
    \caption{
    Time evolution of the reconnection rate $\eta_{\rm rec}=v_{\rm in}/c$ for different values of $\gamma_{\rm  rad}$, as described in the legend.
%
%
%
%
}
    \label{fig:ratess}
\end{figure}

\section{Results}
\label{sec:4results}

%
%

\begin{figure*}[t]
\centering
    \includegraphics[width=\textwidth,height=0.5\textheight,keepaspectratio]{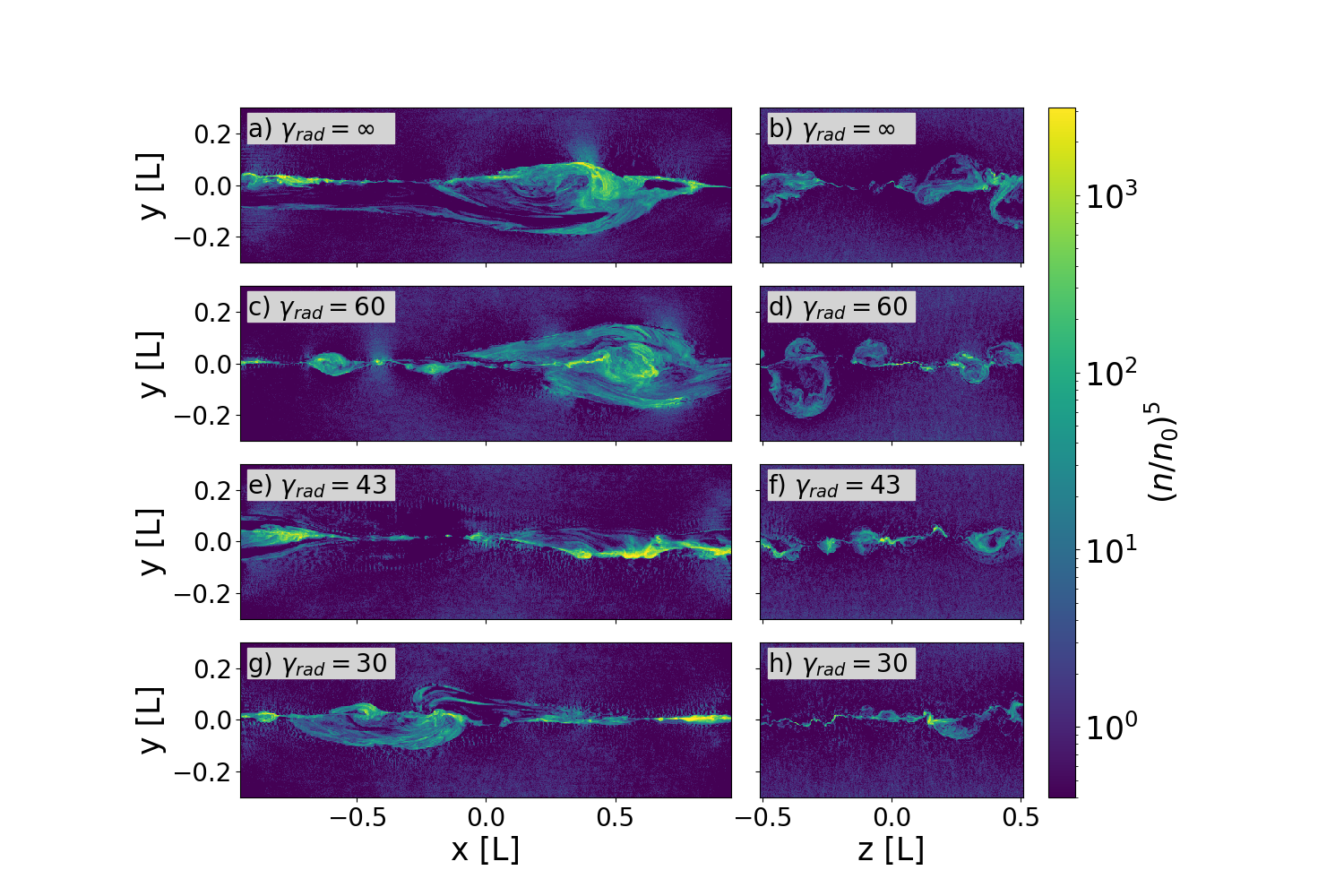}
    \caption{Representative cross sections of the density structure of the reconnection layer, taken at $t= 7.9\, L/c$, for different $\gamma_{\rm  rad}$, as marked on the panels. The density $n$ is normalized to the initial upstream density $n_0$. To enhance contrast, we show $(n/n_0)^5$. Left: Cross sections taken at $z=0$ in the $x-y$ plane, with the inflow direction on the vertical axis and the outflow direction on the horizontal axis. Right: Cross sections taken at $x=0$ in the $y-z$ plane, with the inflow direction on the vertical axis and the guide-field direction on the horizontal axis. 
    }
    \label{fig:denss}
\end{figure*}

  \subsection{Reconnection Rate and Layer Structure}
  \label{sec:41reconnectionrate}
Reconnection is initiated by reducing by hand the particle pressure near the center of the simulation box, generating two fronts that propagate outward along the outflow direction. The two fronts exit the simulation domain after approximately $2\, L/c$, where $L$ is the half-length of the box in the outflow direction. The system then settles into a statistical steady state after $t \gtrsim 4\,L/c$. 

As shown in \figg{ratess}, the reconnection rate clearly reflects these early phases of evolution, with a peak at $t\sim2 \, L/c$ corresponding to the time when the two fronts exit the open boundaries. Here, the reconnection rate is measured as the average inflow velocity (i.e., along $y$) in a strip that extends in $x$ along the whole domain and in $y$ in the range $-0.3\, L<y<-0.15\, L$. We find that the reconnection rate in steady state has no systematic dependence on $\gammarad$ and settles around $\eta_{\rm rec}=v_{\rm in}/c\simeq 0.05$, the same value as in the uncooled case. This provides a useful testbed for comparing our cooled-electron results with the acceleration model of \citet{Zhang_23ads}, which was developed from uncooled runs.


As shown in \figg{denss}, the reconnection layer’s structure and the thickness of the largest plasmoids change little with the level of cooling strength. Uncooled and weakly cooled runs tend to host slightly larger plasmoids, while the layer appears somewhat thinner for stronger cooling; yet even at $\gamma_{\rm rad}=30$ a sizable plasmoid persists. Some differences between the different cases simply reflect the statistical nature of the system.

This contrasts with \citet{Chernoglazov_23ads}, where strong cooling of both species yielded a markedly altered layer with smaller, more compressed plasmoids. Overall, in our case the morphology in \figg{denss} remains broadly consistent across all the cooling cases we explored. We conclude that the fundamental layer physics is largely unchanged among the runs presented here, making them a suitable testbed for the model by \citet{Zhang_23ads}, which was originally developed for uncooled plasmas.

\begin{figure}[h!tbp]
\centering
    \includegraphics[width=\columnwidth]{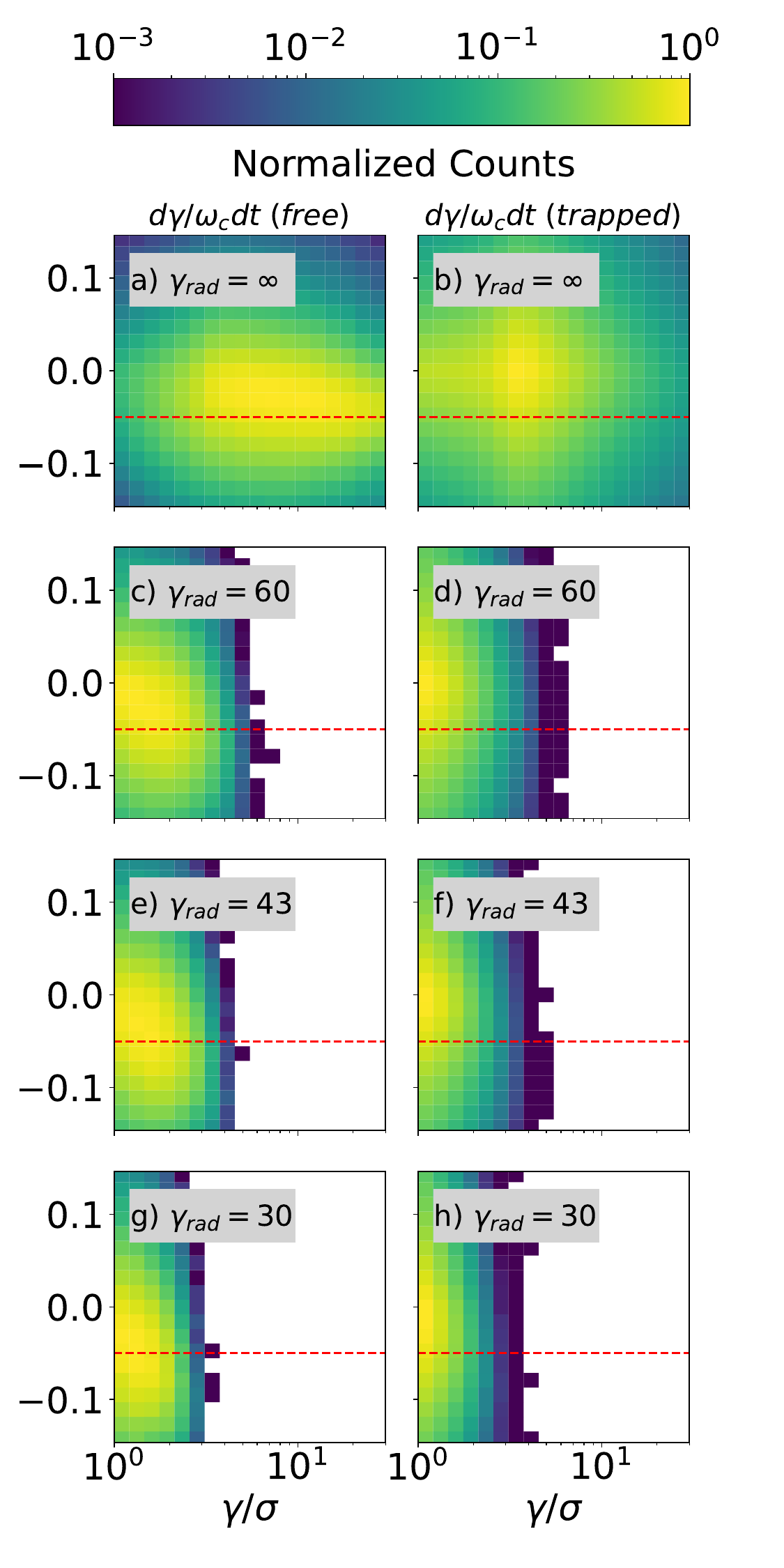}
    \caption{
2D histograms of the acceleration rate $\dot{\gamma}_{\rm acc}/\omega_{\rm c}$ for free electrons (left) and trapped electrons (right), for different $\gammarad$, as marked in the panels. The Lorentz factor $\gamma$ on the horizontal axis is the instantaneous value. The histograms are normalized to their respective maxima, and colors span the range $[10^{-3},1]$ in logarithmic increments. Horizontal dashed red lines 
show the optimal acceleration rate $\dot{\gamma}_{\rm acc}/\omega_{\rm c}\simeq \eta_{\rm rec}\simeq 0.05$.
%
 }
    \label{fig:anisotropp2}
\end{figure}

\begin{figure}[h!tbp]
\centering
    \includegraphics[width=\columnwidth]{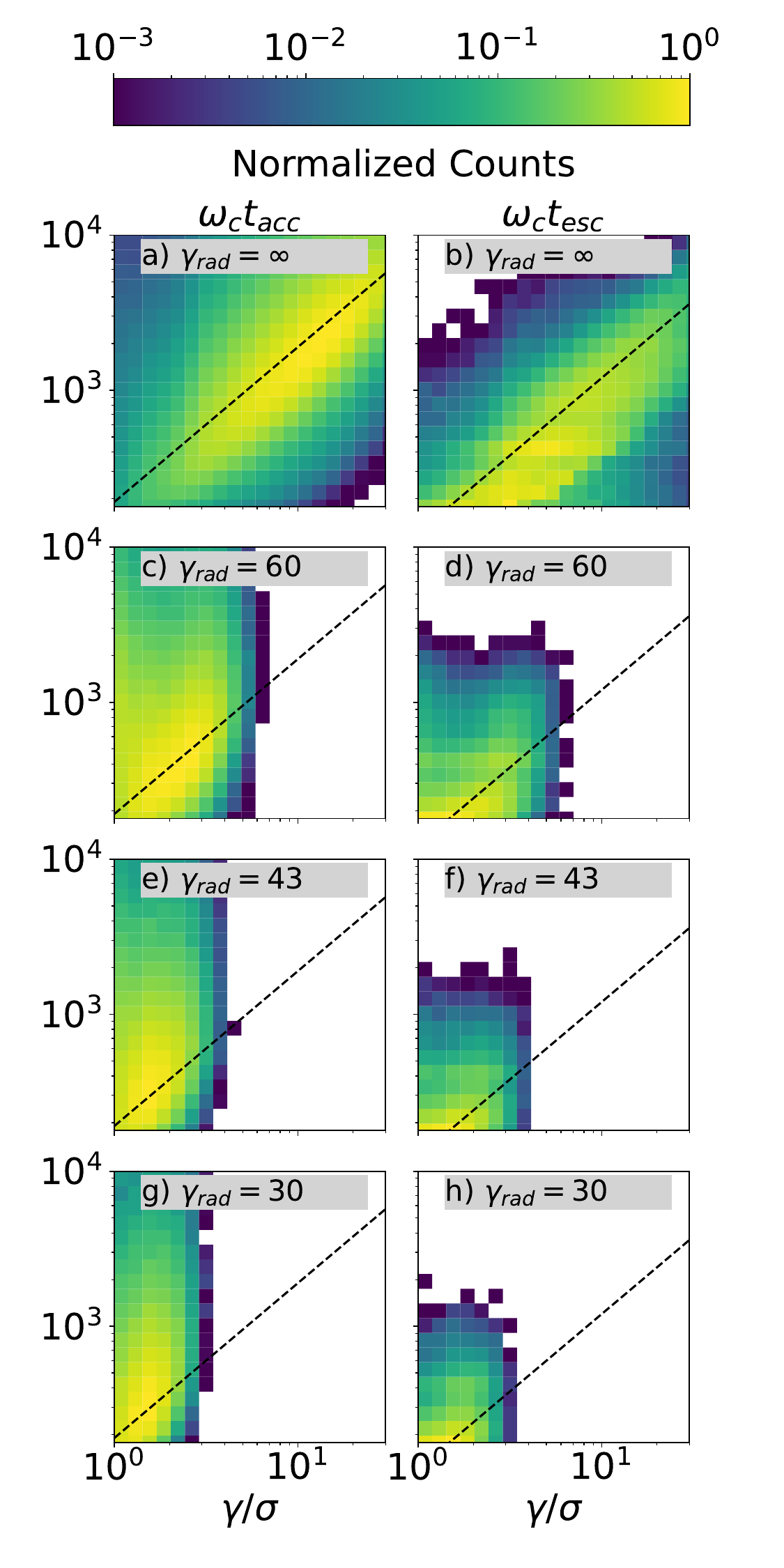}
    \caption{
2D histograms of acceleration time $t_{\rm acc}=\gamma/\dot{\gamma}_{\rm acc}$ (left) and escape time $t_{\rm esc}$ (right) of free electrons, for different $\gammarad$, as marked in the panels. The Lorentz factor $\gamma$ on the horizontal axis is the instantaneous value on the left, while it is the value at the end of the free phase on the right. The histograms are normalized to their respective maxima, and colors span the range $[10^{-3},1]$ in logarithmic increments. Dashed lines are given by $\omega_{\rm c}t_{\rm acc} = 19 \,\gamma$ in the left column, and $\omega_{\rm c}t_{\rm esc} = 13 \,\gamma$ in the right column.
%
}
    \label{fig:tacctescc}
\end{figure}

%


\begin{figure}[h!tbp]
\centering  
\includegraphics[width=0.97\columnwidth]{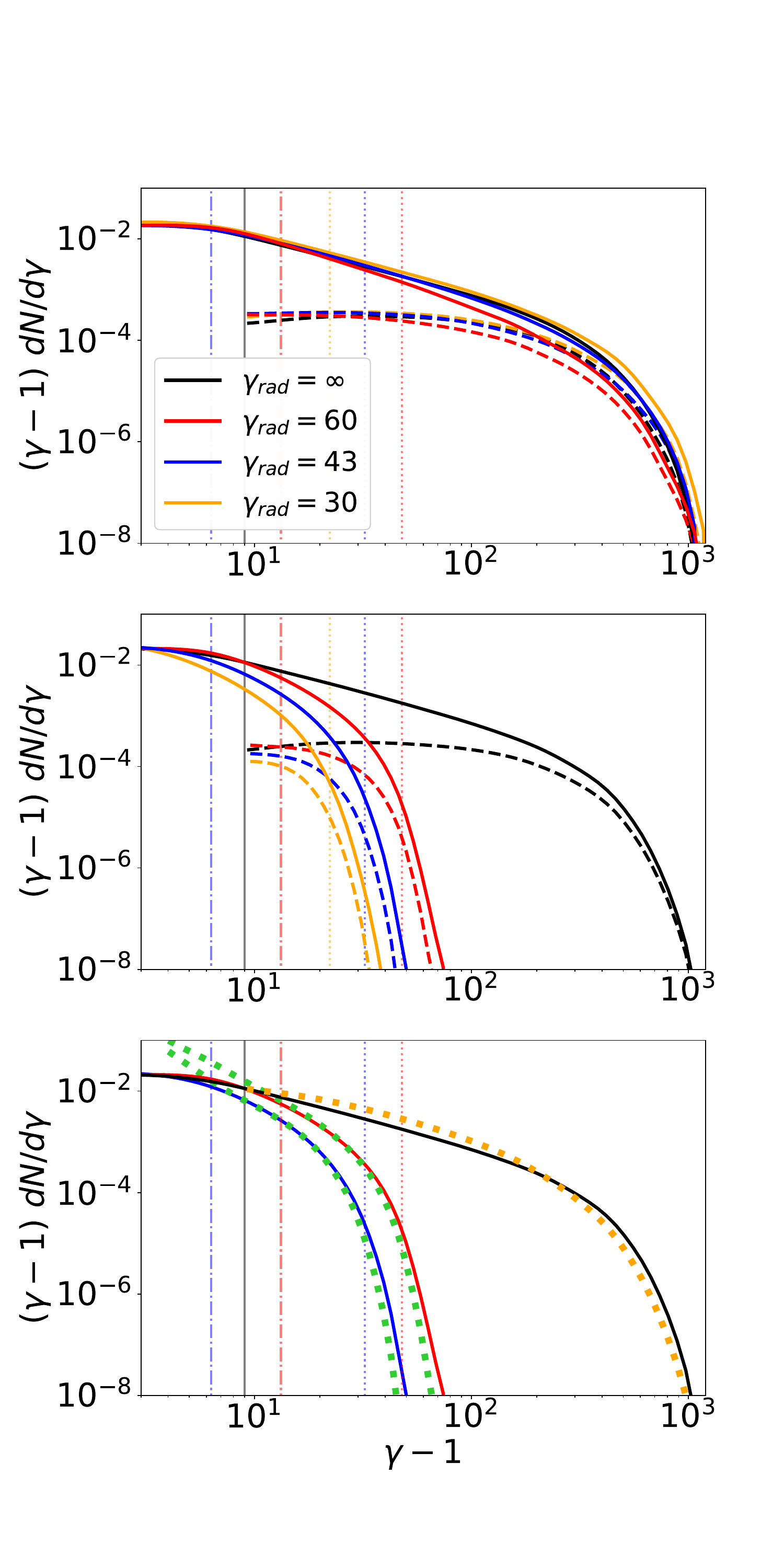}
    \caption{Particle energy spectra averaged during the steady state, $t\gtrsim 4\, L/c$. We show the total spectra (solid; including free and trapped particles) and the spectra of free particles (dashed; shown only for $\gamma>\sigma$). Since free particles are always a minority (apart from the upper spectral cutoff, where they contribute as much as the trapped population), total spectra are nearly the same as the spectra of trapped particles. Top: total and free spectra of positrons. Middle: total and free spectra of electrons. Bottom: total spectra of electrons, and predicted shape of the total spectra based on the measured free spectra and the analytical scalings in \eq{13nocooldiff} (for the uncooled case, yellow dotted curve) or \eq{14cooldiff} (for the cooled cases, green dotted curves). In each panel, the vertical gray line is $\gamma=\sigma$, dotted lines show $\gammarad$, and dot-dashed lines show $\gammacool$.
    %
    %
    %
    }
    \label{fig:spectraa}
\end{figure}

\subsection{Particle Spectrum}
\label{sec:43particlespectrum}
In this subsection, we first characterize the properties of free and trapped electrons, by tracking a large sample of particles. For the uncooled case, we track 0.5 million electrons whose maximum Lorentz factor exceeds $3\,\sigma$; for each of the cooled cases, we track 0.5 million electrons whose maximum Lorentz factor exceeds $\sigma$.
We then use our findings to interpret their energy spectra. 

\begin{figure}[h!tbp]
\centering
    \includegraphics[width=\columnwidth]{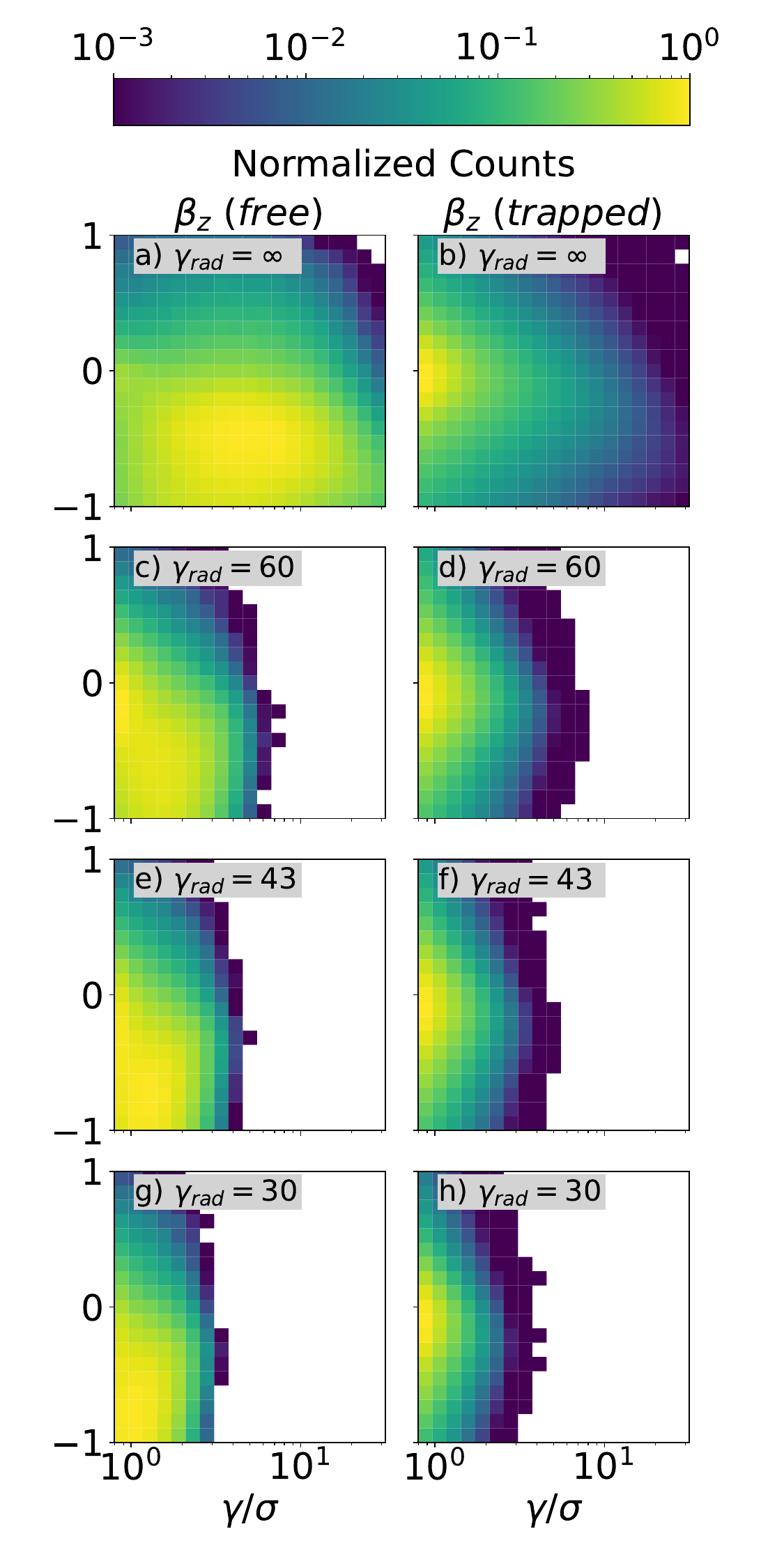}
    \caption{
2D histograms of $\beta_z$ (i.e., the velocity component along the reconnection electric field), for free (left) and trapped (right) electrons and different $\gammarad$, as marked on the panels.
    The Lorentz factor $\gamma$ on the horizontal axis is the instantaneous value. The histograms are normalized to their respective maxima, and colors span the range $[10^{-3},1]$ in logarithmic increments.
    }
    \label{fig:anisotropp}
\end{figure}

To distinguish between upstream (pre-reconnection) and downstream (post-reconnection) regions, we define a ``mixing'' factor \citep{rowan_17,ball_18,sironi_beloborodov_20} $M \equiv 1-2|n_{\rm top}/n - 1/2|$, where $n_{\rm top}$ is the density of particles that started from $y>0$, while $n$ is the total density. We label $M<M_{\rm crit}=0.3$ as  upstream, and $M\geq0.3$ as  downstream, a choice we motivated in \citet{zhang_sironi_21}. If a particle at a given time is in region with $M<0.3$ it is identified as free, otherwise as trapped. We use this criterion, which we shall call ``instantaneous,'' to measure point-wise quantities, e.g., the acceleration rate $\dot{\gamma}_{\rm acc}/\omega_{\rm c}$ in \figg{anisotropp2} or the particle dimensionless $z$ velocity $\beta_z$ in \figg{anisotropp}, and for all the particle energy spectra.

For non-instantaneous quantities, e.g., the acceleration time $t_{\rm acc}$ and the escape time $t_{\rm esc}$ of free particles in \figg{anisotropp2}, it is convenient to employ a more robust strategy that better retains continuity over time.
For each time $t$ along a particle trajectory, we calculate the median of $M$ between $t - t_{\rm L}/2$ and $t + t_{\rm L}/2$, where $t_{\rm L}=2 \pi \gamma \, \omcm$ is the gyration time for a particle with Lorentz factor $\gamma$. If more than 50\% of the median values calculated from $t - t_{\rm L}/2$ to $t + t_{\rm L}/2$ are smaller than $M_{\rm crit}=0.3$, the particle is identified as free at time $t$, otherwise as trapped. We shall call this the ``median'' criterion, which is used for the escape timescale of free particles in \figg{tacctescc}.

In \figg{anisotropp2}, we plot 2D histograms of the electron instantaneous acceleration rate $\dot{\gamma}_{\rm acc}/\omega_{\rm c}$ as a function of $\gamma$, for both free (left) and trapped (right) electrons. The top row shows the uncooled case, and the strength of cooling losses increases towards the bottom row, as marked on the panels. As expected, the horizontal range of the histograms shrinks for stronger losses, since particles can only be accelerated up to $\gammarad$. Other than this effect, the behavior is largely independent from $\gammarad$: the acceleration rate of free particles tends to cluster around the optimal rate $\dot{\gamma}_{\rm acc}/\omega_{\rm c}\simeq \eta_{\rm rec}\simeq 0.05$ (indicated by the horizontal red dashed lines), especially at high energies; in contrast, the histograms of trapped particles are roughly symmetric around $\dot{\gamma}_{\rm acc}/\omega_{\rm c}=0$. Regardless of the strength of cooling losses, this supports one of the main assumptions of the model by \citet{Zhang_23ads}, i.e., that free particles consistently accelerate at nearly the maximal rate, while the trapped phase does not lead to systematic energization.

We then proceed in \figg{tacctescc} to a characterization of the timescales of free electrons, i.e., their acceleration time $\tacc=\gamma/\dot{\gamma}_{\rm acc}$ (left column) and their escape time $\tesc$ from the region of active acceleration (right column).  The escape time $\tesc$ is the duration of each free phase (potentially more than one, for a given particle). We find that, regardless of the strength of cooling losses, both timescales vary linearly with $\gamma$, and the coefficient of proportionality is independent from $\gammarad$. Based on the arguments in Section \ref{sec:2modelstuff}, we would then expect the slope $s_{\rm free}=\tacc/\tesc$ of the free electron spectrum to be the same regardless of the cooling strength. This is supported by the middle panel in \figg{spectraa}, which shows that the spectra of free electrons (dashed colored lines) at $\gamma\ll \gammarad$ are consistent with a power-law scaling $dN_{\rm free}/d\gamma\propto \gamma^{-1}$, as in the uncooled case (black dashed curve). For completeness, we also present the positron spectra (top panel) for the uncooled case (in black) and the cooled cases (in color), showing that positrons consistently display free and trapped spectra identical to the case in which we neglect cooling losses altogether. This supports the notion that the layer dynamics is insensitive to the level of electron cooling we adopt.

We finally discuss the spectra of trapped electrons. In the middle panel of \figg{spectraa}, comparison between the spectra of free (dashed lines) and trapped (solid lines)\footnote{In reality, solid lines show the total spectra, including free and trapped electrons. Since free particles are always a minority (apart from the upper spectral cutoff), total spectra are nearly the same as the spectra of trapped particles.} electrons shows that the two contributions are roughly comparable at the upper cutoff, for both the uncooled case (in black), where the cutoff is at $\gamma_{\rm max}$, as well as for the cooled cases (in color), where the cutoff is controlled by $\gammarad$ (dotted vertical colored lines). This agrees with the expectations presented in Section \ref{sec:2modelstuff}. At high energies, beyond the cooling break $\gammacool$ (dot-dashed vertical colored lines in \figg{spectraa}), the trapped spectra of cooled electrons are appreciably steeper than in the uncooled case. This conclusion is further supported by the bottom panel of \figg{spectraa}. In the uncooled case, the yellow dotted curve shows $\gamma^{-1}dN_{\rm free}/d\gamma$. Based on the arguments in Section \ref{sec:2modelstuff} and following \citet{Zhang_23ads}, we expect that in the uncooled case the trapped spectrum is steeper than the free spectrum by one power of  $(\gamma-1)$, so the yellow dotted line should overlap with the solid black line, as indeed confirmed by the plot. In contrast, for the cooled cases, we expect that for $\gamma>\gammacool$ (the cooling break $\gammacool$ is shown by the dot-dashed vertical colored lines in \figg{spectraa}), the spectrum of trapped electrons should scale as the green dotted lines, which show $(\gamma/\gammarad)^{-2}dN_{\rm free}/d\gamma$. The agreement between the green dotted lines and the colored lines at $\gamma>\gammacool$ then supports the analytical arguments in Section \ref{sec:2modelstuff}, i.e., that the total particle spectra in the range $\max[\sigma,\gammacool]\ll \gamma\ll \gammarad$ scale as $dN/d\gamma\propto \gamma^{-3}$.

We conclude this section by discussing the momentum anisotropy of free and trapped electrons, and in particular their $\beta_z$ velocity. This is the most important component, given that the reconnection electric field is along the $+\hat{z}$ direction, so the electron acceleration rate is expected to be $\dot{\gamma}_{\rm acc}\simeq- \eta_{\rm rec} \beta_z\omega_{\rm c}$. As shown in \figg{anisotropp2}, the acceleration rate of free particles approaches the optimal limit of $\eta_{\rm rec}\omega_{\rm c}$, suggesting that they move with $\beta_z\simeq -1$. This is supported by \figg{anisotropp}, which demonstrates that free electrons (left column) tend to move nearly along the $-\hat{z}$
direction, i.e., their velocity has  optimal orientation for rapid acceleration. In contrast, trapped electrons (right column) have $\beta_z$ symmetrically distributed around zero, which is consistent with the lack of systematic energization seen in the right column of the acceleration rate plot in \figg{anisotropp2}.

\section{Astrophysical Implications}
\label{sec:5discussion}
For the analytical model by \citet{Zhang_23ads} 
to be applicable, the characteristic particle energy scales must satisfy the condition
$\gamma_{\rm cool} < \gamma_{\rm rad} < \gamma_{\rm max}$. When this order is satisfied, the model accurately describes the particle spectrum above the injection energy ($\gamma>\sigma$). In this Section, we show that this condition holds for a wide range of parameters expected in the dissipation region of GRB and blazar jets, assuming that magnetic reconnection powers the observed electromagnetic emission.

So far, the focus of this study has been on plasmas consisting of oppositely charged species of equal mass.
In astrophysical sources, such as blazars and GRBs, the presence of ions (mostly protons) and electrons is expected (with the possible addition of positrons). 
In the presence of multiple particle species, one can define the magnetization for each of them, by normalizing the magnetic enthalpy density to the rest-mass energy density of that species. The electron magnetization, $\sigma_{ e}=B_0^2/4\pi n_0m_ec^2$, which characterizes the energy available per electron in the upstream, is of primary relevance when it comes to modeling the electromagnetic radiation from these sources. Assuming a pure electron-proton plasma, the electron magnetization is $\sigma_{e} \simeq \sigma\, m_{p}/m_{e}$, where $m_p$ and $m_e$ are the masses of protons and electrons, respectively. In reconnection regions of relativistic jets, the overall magnetization  $\sigma$ may be estimated to be of the order of a few (roughly between 1 and 10) 
\citep{10.1093/mnras/stab008}. For instance, in  striped jet models, 
the region of maximum dissipation occurs for $\sigma$ of order unity \citep{10.1093/mnras/stz082}. 
For electrons, $\sigma_{e}$ is therefore a factor of $m_{p}/m_{e}$ larger, implying a rough estimate for $\sigma_{e}$ of around $10^{3}$ to $10^4$.

Assuming a Poynting-flux dominated jet, the Poynting luminosity is comparable to the jet luminosity. This implies that the comoving magnetic field is
\begin{equation} \label{eq:20whatsB}
B_{\rm co}=\frac{\mathcal{L}_{\rm iso}^{1/2}}{c^{1/2}R \Gamma},
\end{equation}
where $\Gamma$ is the bulk Lorentz factor, $R$ is the distance to the central engine, and $\mathcal{L}_{\rm iso}$ is the isotropic jet luminosity.
One can now express the energy scales $\gamma_{\rm  cool}$, $\gamma_{\rm  rad}$, and $\gamma_{\rm  max}$ in terms of quantities such as the bulk Lorentz factor $\Gamma$, 
distance from central engine $R$, size of reconnection region $L$, and isotropic jet luminosity $\mathcal{L}_{\rm iso}$, assuming that synchrotron emission dominates the cooling losses.


\subsection{Blazar Jets}
\label{sec:51blazarjets}

For reconnection regions in blazar jet, we adopt as fiducial reference values $R \simeq 10^{17}$ cm, $\Gamma \simeq 10$, and $L \simeq 10^{16}$ cm and isotropic jet luminosity $\mathcal{L}_{\rm iso} \simeq 10^{48}$ erg/s \citep{10.1093/mnras/stab008}.
 Thus, using the corresponding quantities $L_{\rm 16}$, $\mathcal{L}_{\rm 48}$, $R_{\rm 17}$, and $\Gamma_{\rm 1}$ we have for $\gamma_{\rm  rad}$, $\gamma_{\rm  cool}$, $\gamma_{\rm  max}$:
\begin{equation} \label{eq:22gammacoolB}
    \gamma_{\rm  cool} = 3.5 \times 10^{1} R_{\rm 17}^2 \Gamma_{\rm 1}^2 / \mathcal{L}_{\rm 48} L_{\rm 16},
\end{equation}
\begin{equation} \label{eq:23gammaradB}
	\gamma_{\rm  rad} = 8.4 \times 10^{6}  R_{\rm 17}^{0.5} \Gamma_{\rm 1}^{0.5} /\mathcal{L}_{\rm 48}^{0.25},
\end{equation}
\begin{equation} \label{eq:24gammamaxB}
	\gamma_{\rm  max} = 2.0 \times 10^{12} \mathcal{L}_{\rm 48}^{0.5} L_{\rm 16}/ \Gamma_{\rm 1} R_{\rm 17},
\end{equation}
where we have set the reconnection rate $\eta_{\rm rec} = 0.06$.
With the possible exception of very under-luminous systems, the condition
$\gamma_{\rm cool} < \gamma_{\rm rad} < \gamma_{\rm max}$ holds.

For blazar jets, $\sigma_{e} \simeq 10^3$, so depending on the exact values of $R_{\rm 17}$, $\Gamma_{\rm 1}$, 
$\mathcal{L}_{\rm 48}$, and $L_{\rm 16}$ in \eq{22gammacoolB}, slow cooling could apply. This is likely the case for low luminosity jets; for very powerful blazars such as flat-spectrum radio quasars, the stronger ambient radiation and magnetic fields would lead to $\gamma_{\rm cool}<\sigma_{e}$, i.e., fast cooling conditions.

\subsection{GRB Prompt Emission}
\label{sec:52grb}

Relativistic magnetic reconnection has been proposed as a mechanism to power the GRB prompt emission \citep{spruit_01,lyutikov_03}. 
Typical GRB jet luminosities are on the order of $\mathcal{L}_{\rm iso} \simeq 10^{52}$ erg/s. Though very model dependent, we assume a fiducial distance from the central engine where  magnetic dissipation takes place of $R \simeq 10^{15}$ cm \citep{giannios_05}. 
The bulk Lorentz factor of the jet can be in the range $100<\Gamma<1000$; as a reference, we take it to be $\Gamma \simeq 10^{2.5}$. For the size of the reconnection region in GRBs, we use the causality length, i.e., $L = R/\Gamma$. 
The characteristic electron Lorentz factors are therefore:
\begin{equation} \label{eq:25gammacoolGRB}
    \gamma_{\rm  cool} = 1.1 ~ R_{\rm 15} \Gamma_{\rm 2.5}^3 / \mathcal{L}_{\rm 52},
\end{equation}
\begin{equation} \label{eq:26gammaradGRB}
	\gamma_{\rm  rad} = 4.7 \times 10^{5}  R_{\rm 15}^{0.5} \Gamma_{\rm 2.5}^{0.5} /\mathcal{L}_{\rm 52}^{0.25},
\end{equation}
\begin{equation} \label{eq:27gammamaxGRB}
	\gamma_{\rm  max} = 2.0 \times 10^{11} \mathcal{L}_{\rm 52}^{0.5}/ \Gamma_{\rm 2.5}^{2} .
\end{equation}
As compared to the dissipation regions in blazar jets, the cooling for GRBs is faster (lower $\gamma_{\rm cool}$), stronger (lower $\gamma_{\rm rad}$) and the maximum attainable particle energy is lower (lower $\gamma_{\rm rad}$ and $\gamma_{\rm max}$).


For GRB jets, $\sigma_e$ may be roughly the same as in blazar jets \citep{sobacchi_21b}. However, the approximate value of $\gamma_{\rm cool}$ is three orders of magnitude lower 
than in blazars, so fast cooling is more likely for typical GRB prompt emission regions. The  condition $\gamma_{\rm cool} < \gamma_{\rm rad} < \gamma_{\rm max}$ for the applicability of the model is 
easily satisfied.

\section{Conclusion}
\label{sec:6conclusion}
We have performed a suite of 3D PIC simulations of relativistic reconnection including inverse Compton  losses in the weakly cooled regime in which the radiation-reaction-limited Lorentz factor $\gamma_{\rm rad}$ greatly exceeds the magnetization $\sigma$. We have shown that electron cooling losses do not appreciably alter the reconnection rate, the structure of the layer, and the physics of particle acceleration in the free phase, so the spectrum of free electrons is $dN_{\rm free}/d\gamma\propto \gamma^{-1}$, as in the uncooled case. The spectrum of trapped electrons above the cooling break $\gamma_{\rm cool}$ (in the range $\max[\sigma,\gamma_{\rm cool}]<\gamma<\gamma_{\rm rad}$) is $dN/d\gamma\propto \gamma^{-3}$, steeper than the scaling  $dN/d\gamma\propto \gamma^{-2}$ of uncooled simulations. This confirms that no significant particle energization occurs during the trapped phase, which was one of the fundamental assumptions of the analytical model by \citet{Zhang_23ads}. Our results imply that radiative emission models of reconnection-powered astrophysical sources should employ a two-zone structure, that differentiates between free, rapidly accelerating particles and trapped, passively cooling particles.

The power-law index of the particle spectra, while not directly observable, is an important ingredient when modeling observed photon spectra from objects such as AGN jets \citep{zhang_sironi_21, uhecr_comp_spec, openquestionsuhecrs, brightfermiblazars, thepowerofblazarjets}. In our scenario, the majority of the emitting particles are trapped, so the dominant emission should originate from an electron distribution  
$dN/d\gamma\propto \gamma^{-2}$ steepening to $dN/d\gamma\propto \gamma^{-3}$ above the cooling break. However, free particles become more abundant close to the maximum Lorentz factor $\gamma_{\rm rad}$, which can result in unique observational signatures. For example, free electrons are preferentially beamed along the reconnection electric field. 
Since their IC and synchrotron emission is tightly beamed around their direction of motion, this could possibly result in fast time variability \citep[][]{annurev_DLU}. This could serve as one of the observational diagnostics for testing this model.

We conclude with caveats and potential directions for future investigation. First, the presence of free particles emerges only at $\gamma>\sigma$, so neither this work nor the model by \citet{Zhang_23ads} are applicable to cases of strong cooling, where $\gammarad<\sigma$. Second, the separation between $\gammacool$ and $\gammarad$ in our PIC simulations is much smaller than in realistic astrophysical systems; future work should attempt at broadening the dynamic range, possibly leveraging on the two-zone acceleration picture elucidated in this work. Finally, this work should be extended to the case of electron-proton and pair-proton plasma (with realistic mass ratio) and to reconnection with a significant guide field component.

\acknowledgements
I. B. would like to thank Hao Zhang, Jorge Cortes, Aaron Tran, and Han Zhu for helpful advice. I. B. and D. G. acknowledge support from the NSF AST-2308090 and AST-2510569 grants. 
L.S. acknowledges support from DoE Early Career Award DE-SC0023015, NASA ATP 80NSSC24K1238, NASA ATP 80NSSC24K1826, and NSF AST-2307202. This research was facilitated by the Multimessenger Plasma Physics Center (MPPC) NSF grant PHY-2206609 to L.S., and by a grant from the Simons Foundation (MP-SCMPS-0000147, to L.S.).
%



\appendix

\section{Additional Simulations} \label{appendix:b}
\begin{figure}[htp]
\centering
    \includegraphics[width=\columnwidth]
    {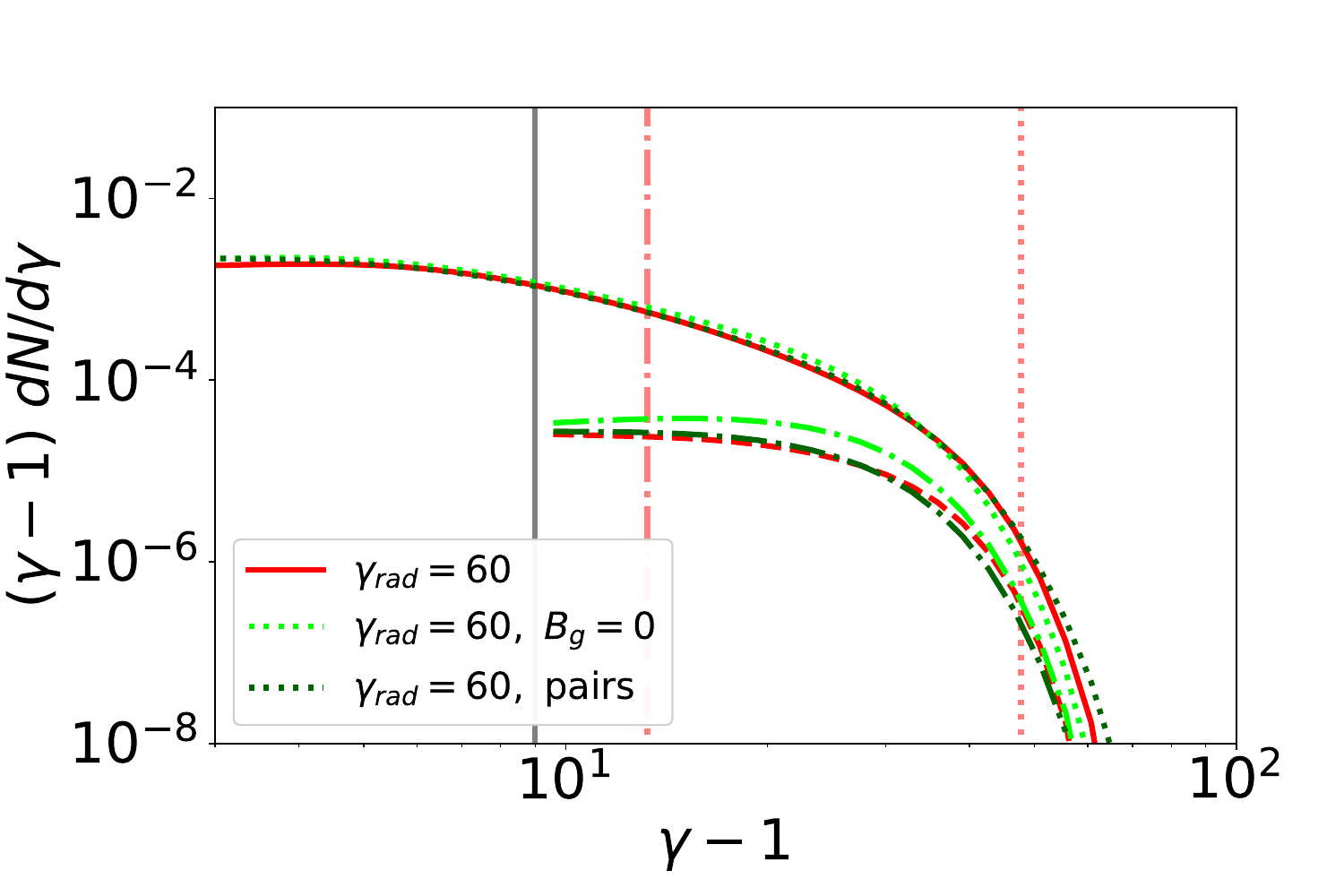}
    \caption{
Electron energy spectra averaged during the steady state, $t\gtrsim 4\, L/c$. We show the fiducial run with $\gamma_{\rm  rad}=60$ in red (solid for all electrons, dashed for free electrons). A run with the same physical and numerical parameters, but where we cool both species, is shown in dark green (dotted for all electrons, dot-dashed for free electrons). A run with the same parameters as our fiducial one but with $B_{\rm g}=0$ instead of $B_{\rm g}=0.1\,B_0$ is shown in light green (dotted for all electrons, dot-dashed for free electrons). Free electron spectra are shown only for $\gamma>\sigma$. The vertical gray line is $\gamma=\sigma$, the dotted  vertical red line is $\gammarad$, and the dot-dashed vertical red line is $\gammacool$.
}
    \label{fig:pairBg60}
\end{figure}

In addition to the simulations presented in the main body of the paper, we corroborate our results with two additional simulations, whose free and trapped electron spectra are shown in \figg{pairBg60}. One of them is performed without any guide field, and we cool only electrons. The other uses the fiducial guide field of $B_{\rm g}=0.1 \, B_0$ and cools both species.
Otherwise, the two additional simulations adopt the same physical and numerical parameters as the runs discussed in the main paper. In particular, we use $\gammarad=60$. As shown in  \figg{pairBg60}, the spectra of free and trapped electrons are the same as in the fiducial $\gammarad=60$ run presented in the main body of the paper.

\bibliographystyle{aasjournal}
\bibliography{blob}

\end{document}